\documentclass{article}

\usepackage{arxiv}

\usepackage[utf8]{inputenc} % allow utf-8 input
\DeclareUnicodeCharacter{200A}{!!!FIX ME!!!} 
\usepackage[T1]{fontenc}    % use 8-bit T1 fonts
\usepackage{hyperref}       % hyperlinks
\usepackage{url}            % simple URL typesetting
\usepackage{booktabs}       % professional-quality tables
\usepackage{amsfonts}       % blackboard math symbols
\usepackage{nicefrac}       % compact symbols for 1/2, etc.
\usepackage{microtype}      % microtypography
\usepackage{graphicx}
\usepackage{natbib}
\usepackage{doi}
\usepackage{wrapfig}
\usepackage{caption}
\usepackage{multicol}

\title{Finding Vulnerabilities in Mobile Application APIs: \\A Modular Programmatic Approach}

\author{Nate Haris, Kendree Chen\thanks{Co-Lead Author, Lead Developer}, Ann Song, Benjamin Pou\\[1ex]
\small{Additional Contributors: Abhinav Sood, Andrew Cao, Tiffany Li, Ronit Barman}}
\renewcommand{\headeright}{Research Paper}
\renewcommand{\undertitle}{Research Paper}
\renewcommand{\shorttitle}{}

\hypersetup{
pdftitle={Finding Vulnerabilities in Mobile Application APIs: A Modular Programmatic Approach},
pdfsubject={arXiv.CS},
pdfauthor={N.~Haris, K.~Chen, A.~Song, B.~Pou},
pdfkeywords={API, API Security, API Vulnerability, API Testing, BAC, Endpoint Vulnerability, OWASP Vulnerability},
}

\begin{document}
\maketitle
\begin{abstract}
	Currently, Application Programming Interfaces (APIs) are becoming increasingly popular to facilitate data transfer in a variety of mobile applications. These APIs often process sensitive user information through their endpoints, which are potentially exploitable due to developer misimplementation. In this paper, a custom, modular endpoint vulnerability detection tool was created and implemented to present current statistics on the degree of information leakage in various mobile Android applications. Our endpoint vulnerability detection tool provided an automated approach to API testing, programmatically modifying requests multiple times using specific information attack methods (IAMs) and heuristically analyzing responses for potentially vulnerable endpoints (PVEs). After analysis of API requests in an encompassing range of applications, findings showed that easily exploitable Broken Access Control (BAC) vulnerabilities of varying severity were common in over 50\% of applications. These vulnerabilities ranged from small data leakages due to unintended API use, to full disclosure of sensitive user data, including passwords, names, addresses, and SSNs. This investigation aims to demonstrate the necessity of complete API endpoint security within Android applications, as well as provide an open source example of a modular program which developers could use to test for endpoint vulnerabilities. 
\end{abstract}


\section{Introduction}
\label{sec:Introduction}
\begin{multicols}{2}
APIs serve as a multi-purpose communication medium between client and server. APIs channel every user-sent request to the server, carrying back a response. Without APIs, many applications would not be able to function as it would be impossible to transport the requested and provided information.

Due to their effectiveness and ease of use, APIs are commonly used by application developers. However, APIs become security risks when they are improperly designed or implemented. Hackers can exploit API vulnerabilities to harm servers or cause them to perform unintended actions, which can have dire consequences for developers and their users. Many small applications implementing personal APIs instantiate vulnerabilities in their backend due to the diverse number of potential attack points. Excessively complex or poorly documented APIs can obstruct developers from testing extensively for security flaws. Thus, it is often simple for third parties to gain sensitive information from server-side devices or transfer unauthorized controls in a network. 

API vulnerabilities are characterized based on weak points, commonality, and severity to users and applications. The Open Web Application Security Project (OWASP), one of the leading community organizations of cybersecurity professionals and penetration testers, publishes a standard awareness document for the state of web application security every four years. 

\paragraph{Broken Access Control (BAC).}
Broken access control is a vulnerability allowing users to access resources or perform actions that they should not have authorization for. In 2021, OWASP’s “Top Ten'' most vulnerable web application security risks featured Broken Access Control, most notably the Common Weakness Enumerations CWE-200 and CWE-201 – exposure or insertion of sensitive data by a user – rising four places to the highest position, A01, due to its mass max coverage rate of 94.55\%. As defined by OWASP, Broken Access Control (BAC) is “unauthorized information disclosure, modification, or destruction of all data or performing a business function outside the user's limits.”

Since BACs are listed as a widespread vulnerability among applications tested by OWASP, this paper was dedicated to assessing the effectiveness of potential methods of abusing this insecurity. The free web traffic applications HTTP Toolkit and Insomnia REST were used to analyze requests from a multitude of Android applications. Furthermore, a script to send modified API requests was run on a cross-section of Android applications to analyze susceptibility to programmatic data scraping. After finding numerous exploitable vulnerabilities, one can conclude that proper API security should be prioritized by application developers, with at minimum rigorous testing necessitated in the context of sensitive information. We hope that our efforts and conclusion will raise awareness among developers that user data and API endpoint protection is of utmost importance.  
\end{multicols}

\section{Literature Review}
\label{sec:Literature Review}
\begin{multicols}{2}
Throughout the last twenty years, more and more companies have begun storing data in large-format databases (\cite{9034254}, \cite{Keary_2022}). The entry of data to these databases is facilitated by various Application Programming Interfaces (APIs). APIs allow for easy transfer of data between applications but are not always secure (\cite{10.1007/978-981-16-9229-1_11}). Numerous private corporations have experienced large-scale information leakage due to API vulnerabilities in recent years (\cite{9799221}, \cite{SaltLabs_2021}).  These API endpoint vulnerabilities are becoming increasingly common in applications (\cite{Keary_2022}), posing a massive threat to organizations that rely on APIs to transport data through the web (\cite{8835301}). 

The most prevalent API vulnerability pertains to BACs (\cite{Shkedy_2019}). Data shows that this specific vulnerability is becoming increasingly common, while also presenting critical level security threats in applications (\cite{Keary_2022}, \cite{Wang_2019}). There already exist many commercial tools that can test for this vulnerability (\cite{9721567}, \cite{8536162}), as well as advanced theoretical detection methods (\cite{10.21681}, \cite{266592}, \cite{7163061}), however the costly, complicated, and time-consuming nature of these analysis methods sometimes renders them inaccessible to independent or novice developers (\cite{Korolov_2023}).

Although this presents a challenge for many developers, there also exist open-source methods of assessing API security (\cite{10.14419}, \cite{10.1007/978-981-16-9229-1_11}, \cite{Wang_2019}). Using free API modification programs, one can independently intercept and scrutinize API requests (\cite{HTTPToolKit}, \cite{AndriodStudio}, \cite{InsomniaRest}) and create a programmatic approach where APIs are tested using a custom set of criteria (\cite{10.14419}). However, such programs often lack the flexibility required to test different APIs, or fail to corroborate with alternative testing programs on the status of vulnerabilities in these applications (\cite{7427159}). 

Due to the confusing state of API program testing, programs with simplistic yet modular frameworks are imperative to assisting developers find vulnerabilities (\cite{10.1007/978-981-16-9229-1_11}). This paper aims to improve the state of programmatic API testing by creating a new flexible, expandable, and accessible framework for application developers.
\end{multicols}

\section{Methodology}
\label{sec:Methodology}
To understand and assess the current state of the API vulnerabilities in Android applications, an inclusive range of applications was analyzed. 72 applications were downloaded and tested using a static environment of Android Emulator to reduce systematic differences between tester hardware. After an extensive test of API function calls in each application, contributors utilized HTTP Toolkit to download .har files containing request data from each API. In the case of a request necessitating manual testing, Insomnia REST was used to modify requests. Similarly, case studies or severe vulnerabilities were further analyzed using Insomnia REST. 

The process of sending requests to APIs was automated through a Java program that parsed large quantities of app traffic as .har data. Data was stored in a SQL database with two tables. Table 1 held collected request data and the results of modified requests, while the Table 2 held request headers for each request stored in Table 1. Table 1 recorded request id number, request method, url, request body, response status, type, and body, and a description of the modification that was made. For each header stored in Table 2, header id number, the id number of the request the header belonged to, header name, and value were included [Figure 1]. 
\newpage

\begin{multicols}{2}
\hspace{0pt plus 1fill}\begin{minipage}{0.62\textwidth}
\begin{minipage}{9cm}
\centering
    \includegraphics[width=9cm]{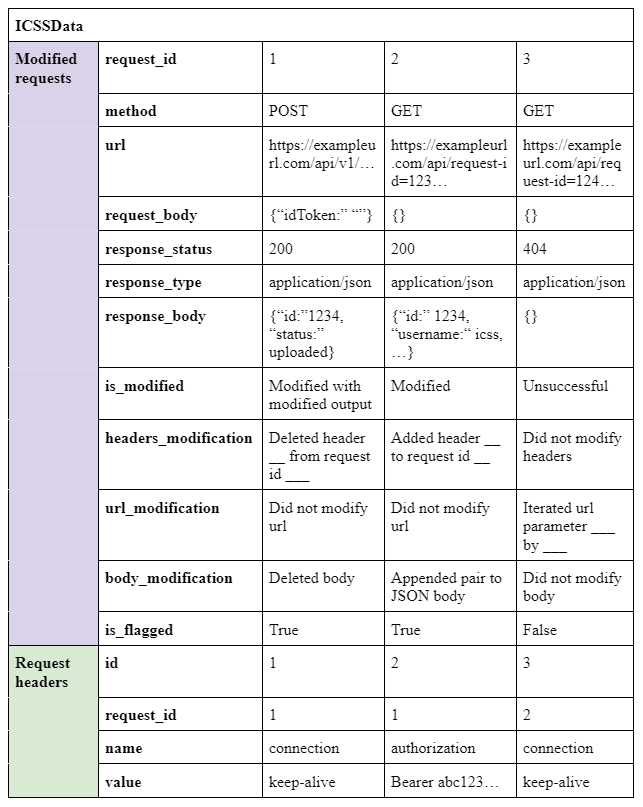}
    \captionsetup{width=.95\textwidth}
    \captionof{figure}{Example layout showing entries of database tables storing information on sent requests and request headers.}
    \label{fig:fig1}
\end{minipage}
\\

\begin{minipage}{9cm}
       \centering
	\includegraphics[width=9cm]{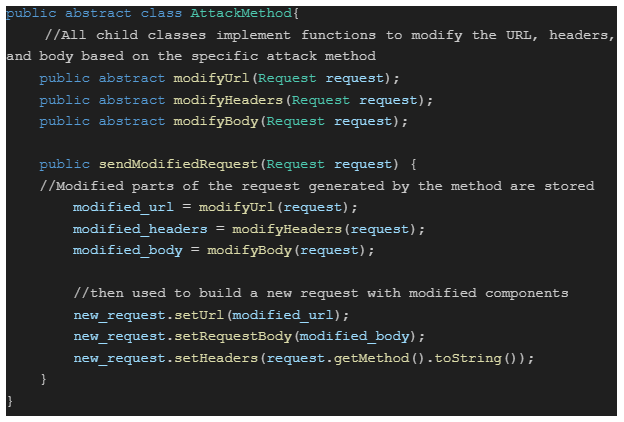}
        \captionsetup{width=1\textwidth}
	\captionof{figure}{Simplistic view of IAM interface implemented by all methods. The relevant modification functions are overridden by each implementation to edit API request body, headers, or URL in a way pertaining to the strategy.}
	\label{fig:fig2}
\end{minipage}

\begin{minipage}{9cm}
    \centering
    \includegraphics[width=9cm]{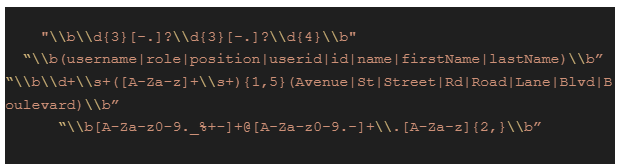}
    \captionsetup{width=1.2\textwidth}
    \captionof{figure}{Regex strings to parse responses for sensitive information.}
    \label{fig:fig3}
\end{minipage}

\end{minipage}
\columnbreak

\hspace{0pt plus 1fill}\begin{minipage}{0.4\textwidth}

\paragraph{Information Attack Method (IAM)} 
Information Attack Method (IAM). Request modifications were made through programmatic generation strategies, which targeted common BAC vulnerabilities such as IDOR parameters to find API endpoints vulnerable to the extraction of sensitive information. In the script, each IAM implemented a parent interface that allowed each strategy to modify different parts of a request to exploit potential vulnerabilities [Figure 2]. This modular approach maintains program scalability. IAMs can be easily added or updated to test varying strategies and test for different types of vulnerabilities within the API. The modifications tested included iterating user ID numbers, removing headers, changing URL parameters, removing request bodies, adding text to headers, and appending JSON data.
\\

\paragraph{Potentially Vulnerable Endpoint (PVE).} 
Responses from modified requests were stored in the first database table and parsed for regex keywords and Levenshtein distance comparison to the original returned data [Figure 3, 4], except for those that returned website CSS, HTML, or JavaScript code or text strings greater than 100,000 characters long; these were recorded and parsed manually for runtime efficiency. Responses with greater than 90\% dissimilarity that contained data-sensitive keywords were flagged as PVEs. Statistics for the success rate, number of dissimilar responses, and number of PVEs generated by each IAM were generated for each app tested [Figure 3]. The script for programmatic IAM testing was run on each of the applications in the collected cross-section to provide statistics about PVE prevalence in the market.
\\

\vspace{20pt}

\begin{minipage}{8cm}
    \centering
    \includegraphics[width=8cm]{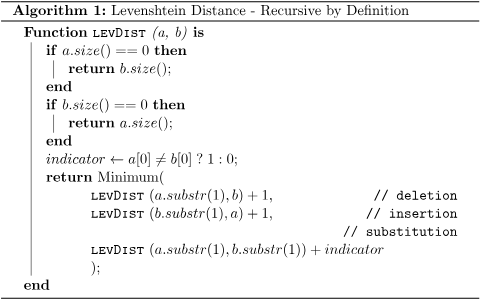}
    \captionsetup{width=1\textwidth}
    \captionof{figure}{Regex strings to parse responses for sensitive information.}
    \label{fig:fig4}
\end{minipage}
\end{minipage}

\end{multicols}
\newpage

\begin{multicols}{2}

\begin{minipage}{.45\textwidth}
\paragraph{Response PVE Heuristics}
Responses with greater than 90\% dissimilarity that contained data-sensitive keywords were flagged as PVEs. Statistics for the success rate, number of dissimilar responses, and number of PVEs generated by each IAM were generated for each app tested [Figure 5]. The script for programmatic IAM testing was run on each of the applications in the collected cross-section to provide statistics about PVE prevalence in the market.
\end{minipage}

\columnbreak

\begin{minipage}{.85\textwidth}
    \begin{minipage}{9cm}
    \centering
    \includegraphics[width=\linewidth]{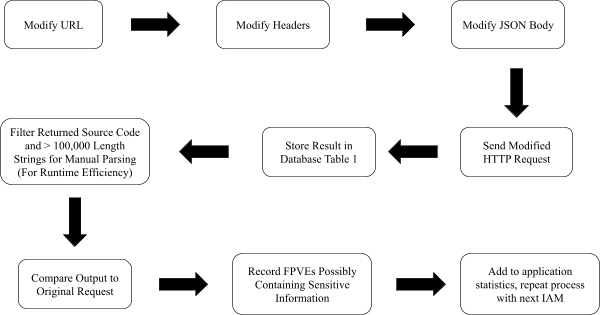} % Adjust the width to match the available space
    \captionsetup{width=1\textwidth}
    \captionof{figure}{Flowchart detailing workflow sequence of programmatic request modification and vulnerability detection.}
    \label{fig:fig5}
    \end{minipage}
\end{minipage}

\end{multicols}

\section{Results}
\begin{multicols}{2}
 72 Android applications were analyzed in total. PVEs were recorded for \textbf{54.3\%} of applications, denoting that they contained potentially sensitive new information that had less than 90\% similarity to the original result [Figure 6].

 Different IAMs varied in the number of PVEs produced [Figure 7], with IAMs manipulating API request URLs producing the most by far. This suggests that many of the detected PVEs were due to insecure direct object references (an example of BAC), as manipulation of references in the parameters of the API request URL generated the greatest number of PVEs. 

Although the majority of analyzed applications contained PVEs, flagging them was susceptible to false positives, such as if a request returned the user’s own data with other dissimilar text. Nevertheless, PVEs efficiently indicated potential attack routes for sensitive information handled by an API. With manual parsing of PVEs, the method undertaken expedited and simplified finding PVEs in applications. Approximately 25\% of PVEs displayed serious exploitable vulnerabilities when manually tested, mainly including the release of sensitive data, demonstrating the potential of programmatic request modification to unearth private information.

Examples of such exploitable vulnerabilities were noticed in numerous applications (mainly consisting of the following CWEs: 200, 201, 285, 359, 538, 540, 862). To better understand some of these critical vulnerabilities found in application analysis, a number of case studies will be included.

\columnbreak
\begin{minipage}{9cm}
    \centering
    \includegraphics[width=9cm]{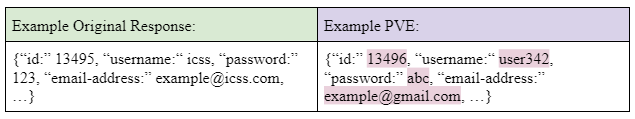}
    \captionsetup{width=1\textwidth}
    \captionof{figure}{Example of a flagged PVE response. Highlighted text signifies response dissimilarities.}
    \label{fig:fig6}
\end{minipage}

\begin{minipage}{9cm}
    \centering
    \includegraphics[width=9cm]{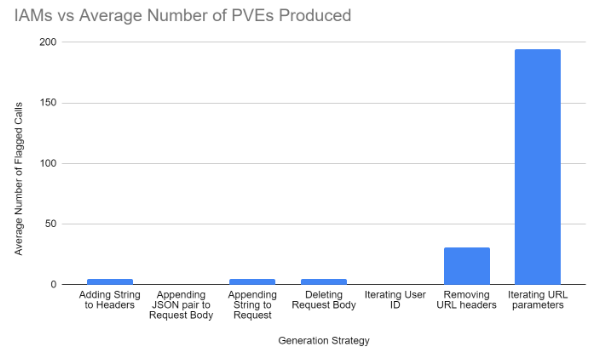}
    \captionsetup{width=1\textwidth}
    \captionof{figure}{Histogram showing average number of flagged calls per IAM}
    \label{fig:fig7}
\end{minipage}
    
\end{multicols}
    
\section{CWE Vulnerabilities 200, 201, 285, 359, and 862 (Case 1).}

In the analysis of 72 various Android applications, vulnerabilities pertaining to CWEs 200, 201, 285, 359, and 862 (information data leakage through insecure objects or authentication) were found to be the most prevalent and most severe. Roughly 20\% of PVEs contained some form of these vulnerabilities. Numerous vulnerable API endpoint calls were identified where, through simple programmatic modifications of API URLs, headers, or authentication functions, a third party could access sensitive user data of various private accounts.

Broken access control, the subgrouping of the aforementioned CWEs, as defined by OWASP, is characterized by a vulnerability allowing a user to access information outside of their specified permissions. In the instances of the Android applications, many vulnerabilities were caused by IDORs, allowing for horizontal privilege escalation. One such possible example of a vulnerable API request could be demonstrated by the following URL, classified as a CWE-359:

\begin{center}
	\url{https://example-site/users/get-info/?user=13495}
\end{center}

In this URL, “user” is an insecure direct object reference and an example of a BAC vulnerability. Modeled after a specific issue detected in a number of Android applications, iterating the “user” parameter could allow one to access users’ data. This is a highly critical privilege escalation vulnerability that could lead to the leakage of sensitive data in the response body [Figure 8].

\begin{center}
\begin{minipage}{11cm}
    \centering
    \includegraphics[width=11cm]{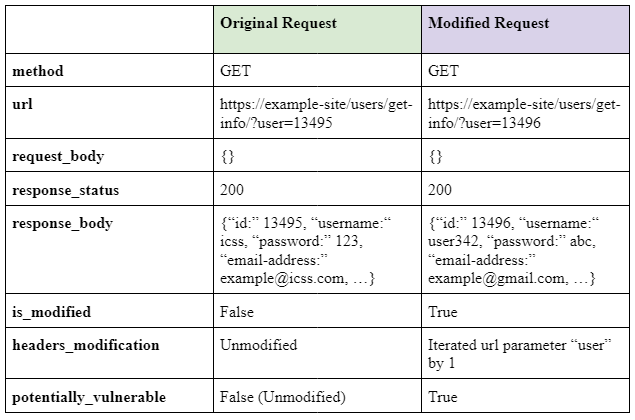}
    \captionsetup{width=1\textwidth}
    \captionof{figure}{Example of original and modified responses of exploitable IDOR vulnerability}
    \label{fig:fig8}
\end{minipage}
\end{center}

The most effective approach to negating this vulnerability is through an implementation of authentication in user API requests. With functional authentication keys, only authorized users will be able to access data, regardless of the user-id parameter. An attempt at accessing private data would result in a 401 error code (unauthorized response).

\subsection{CWE Vulnerabilities 538 and 540 (Case 2).}

\begin{wrapfigure}{r}{8cm}
    \centering
    \includegraphics[width=8cm]{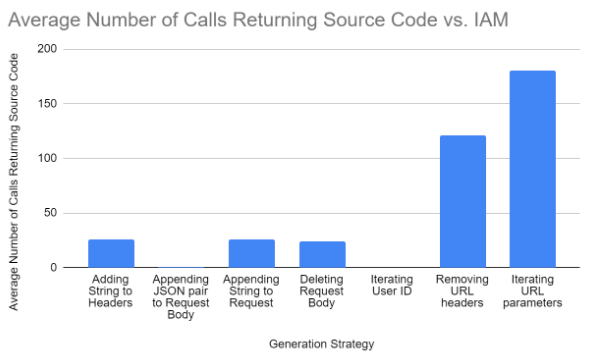}
    \captionsetup{width=0.4\textwidth}
    \caption{Histogram showing the average number of calls returning application source code per IAM.}
    \label{fig:fig9}
\end{wrapfigure}

In some select cases during analysis, the programmatic flagging method and sensitive information regex identifier detected instances of CWE-538 and CWE-540. Both of these vulnerabilities are associated with leakage of sensitive data in source code or external files. In certain API endpoint calls, modification of the URL, header, or body would result in the disclosure of source code or external files containing sensitive data [Figure 9]. IAMs had varying proportions of calls that returned code, with URL modifications returning the most. This supported the claim that URL modification often serves as the entry point for many broken access control vulnerabilities [Figure 7, 9].

Such vulnerability is an example of insecure information management and susceptibility to BAC. Due to the iterative nature of this vulnerability, simple programmatic request modification could reveal sensitive information. A possible vulnerable API request demonstrating this vulnerability is as follows:

\begin{center}
	\url{https://example-site/retrieve-data/?source=ABPQqLF6Vjt6Pzz7lHjE-CRr….}
\end{center}

In this URL, a modification of the source text field can expose an insecure information management vulnerability. With enough iteration of the source text field, previously inaccessible data could be disclosed to an unauthorized user. Although this vulnerability can leak highly sensitive information, it is much more difficult and time-consuming to exploit. Among the numerous applications analyzed, this vulnerability was only discovered a few times.

The most effective approach to negating this vulnerability would be through the implementation of authentication during API requests. In this example, access privileges would be denied to users without an auth key or token. This added security would protect sensitive information, causing unauthenticated API requests would result in a 401 error (unauthorized response). 

\subsection{False Positive Potentially Vulnerable Endpoints (FPPVEs).}
Although flagged PVEs often indicated vulnerabilities, there were some cases where PVEs produced false positives, or API responses which did not exploit or reveal sensitive information (FPPVEs).

FPPVEs occurred due to the criteria for classifying modified API responses as PVEs, which was intentionally generalized to allow for vulnerability detection in a wide range of applications. The purpose of the Levenshtein distance formula and sensitive information regexes was to identify modified requests which contained some form of sensitive data, yet were also different from the original response, as to eliminate classifying a call as a PVE with sensitive information pertaining to the original user. For example, consider an API call which returns the user’s id, authentication, and login credentials. If any modification to this API call similarly returns the original user’s own data, although the information is sensitive, it was previously available to that user, meaning it should not be considered as a vulnerability. The 90\% dissimilarity threshold was established to try to reduce FPPVEs of this type, flagging only unique data, however specific cases still satisfied the conditions for a PVE without containing truly sensitive information. 

A case which caused this type of FPPVE was the exposure of unique and sensitive, but public, information. This case was manually detected in many commercial locater APIs. Imagine the mobile application of an example restaurant. When ordering takeout on such an application, there is a search feature to find only the closest restaurants to the user. This is authorized through an API, but upon modifying these requests, users can extract addresses of previously hidden restaurant locations [Figure 10]. This would result in a response over 90\% dissimilar to the original request, while also satisfying the address regex to detect sensitive information, thus flagging the request as a PVE.

\begin{center}
\begin{minipage}{11cm}
    \centering
    \includegraphics[width=11cm]{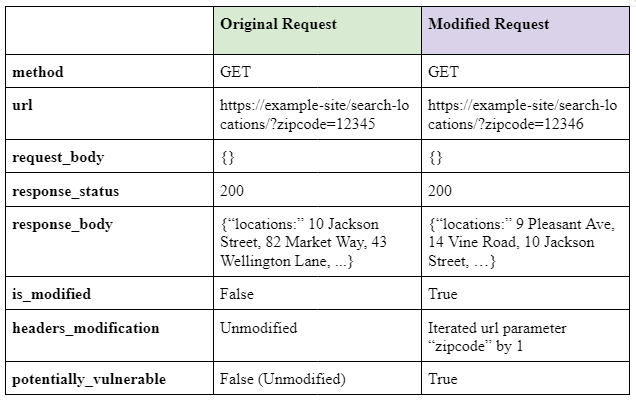}
    \captionsetup{width=1\textwidth}
    \captionof{figure}{Example of false positive potentially vulnerable endpoint (FPPVE)}
    \label{fig:fig10}
\end{minipage}
\end{center}

However, in this case the locations of the example restaurant are publicly available, possible to access through the intended use of the API and a location modification software such as a VPN, so the extracted addresses through the modified request do not leak any sensitive information. This results in a FPPVE where the unique sensitive information revealed was, in reality, not sensitive.

FPPVEs were common among applications with identified PVEs, and thus manual revision was necessary for distinct cases. Upon analysis, PVE’s could easily be sorted into major vulnerabilities or FPPVEs. These FPPVEs could then offer insight on minor vulnerabilities, such as potential attack routes through exploitable API endpoint behavior. FPPVEs might not have directly caused the exposure of truly sensitive data, however they frequently performed actions outside of the intended use of the API endpoint, or accessed data outside the scope of the user’s privileges. Therefore, FPPVEs meaningfully contributed to a further understanding of the limits and minor vulnerabilities persistent in an API endpoint. Through the manual parsing of all PVEs —including FPPVEs— the integrity of an API’s access control protocols could be holistically analyzed and categorized into its minor and major vulnerabilities.

\section{Conclusion}
With the conclusion of this investigation, we hope to reinforce the concerning statistics of the nature of API vulnerabilities, and provide a flexible programmatic approach to API endpoint testing. Vulnerabilities, specifically pertaining to BAC, are becoming increasingly common among Android applications due to improper API implementation. A majority of applications contain some form of minor vulnerability, while a smaller proportion contain a major security flaw pertaining to the storage of sensitive user information. (Our team has notified applications where these vulnerabilities were found, presenting the nature of the flaw and how to properly secure the API). We hope to provide a modular, open-source Java program that may be used as a tool to test for BAC vulnerabilities. Our code, \href{https://github.com/gmsong06/icss-22-23-code}{available on GitHub}, can be used to help identify these specific vulnerabilities quickly and efficiently in mass testing of API function calls.

Due to the data presented on the commonality of these vulnerabilities, we urge application developers to be conscious of the potentially insecure nature of their API endpoints. With the current portfolio of applications and open-source API testing programs, we advise each developer to personalize extensive testing for any potential attack points in their APIs. 

\section{Acknowledgements}
Thanks to Benjamin Zhu and Yiray Wang for assistance in data collection.

\bibliographystyle{unsrtnat}
\bibliography{references} 
%%% Uncomment this line and comment out the ``thebibliography'' section below to use the external .bib file (using bibtex) .

\end{document}